%% file: intel-hexl.tex
\newif\ifiacr
\newif\ifarxiv
\newif\ifacm

\arxivtrue
\acmfalse
\iacrfalse

\ifiacr
\documentclass{llncs}
\usepackage{caption}
\usepackage{cite}
\else

\documentclass[sigconf]{acmart}
\usepackage{balance}
\usepackage{caption}
\fi

\def\BibTeX{{\rm B\kern-.05em{\sc i\kern-.025em b}\kern-.08emT\kern-.1667em\lower.7ex\hbox{E}\kern-.125emX}}

\usepackage{microtype}
\usepackage{graphicx}
\usepackage{tablefootnote}
\usepackage{makecell}
\usepackage{booktabs} 
\usepackage{multirow}
\usepackage{listings}
\usepackage{graphicx}
\usepackage{color}
\usepackage{algorithm}
\usepackage{algorithmicx}
\usepackage{algpseudocode}
\usepackage{tikz}
\usetikzlibrary{arrows}
\usepackage{textcomp} 
\usepackage[htt]{hyphenat} 

\usepackage[T1]{fontenc}
\newcommand{\StateD}[1]{\State{\detokenize{#1}}}

\usepackage{hyperref}

\usepackage{xcolor} 

\usepackage{pifont}
\usepackage{stfloats} 
\usepackage{soul} 
\usepackage{tikz}
\usepackage[normalem]{ulem} 

\usepackage[separate-uncertainty=true,load=abbr.,load=prefixed,decimalsymbol=fullstop,digitsep=comma,sepfour,per-mode=fraction,binary-units=true]{siunitx}
\usepackage{siunitx}
\usepackage{amsmath}
\usepackage{amsfonts}

\usepackage{amssymb}
\usepackage[nointegrals]{wasysym} 
\usepackage{booktabs}

\newunit[unitsep=space]{\imagespersec}{images\per\second}
\newunit[unitsep=space]{\impersec}{im\per\second}
\newunit[unitsep=space]{\msperimage}{\ms\per image}
\newunit[unitsep=space]{\gbperimage}{\giga\byte\per image}
\newunit[unitsep=space]{\mbperimage}{\mega\byte\per image}
\newunit[unitsep=space]{\mbperim}{\mega\byte\per im}

\ifiacr
\else

\fi

\definecolor{mygreen}{rgb}{0,0.6,0}
\definecolor{mygray}{rgb}{0.5,0.5,0.5}
\definecolor{mymauve}{rgb}{0.58,0,0.82}

\ifarxiv
\renewcommand\footnotetextcopyrightpermission[1]{} 
\fi

\ifiacr
\fi

\ifacm
\setcopyright{acmcopyright}
\copyrightyear{2021}
\acmYear{2021}
\acmConference[WAHC'21]{9th Workshop on Encrypted Computing \& Applied Homomorphic Cryptography}{November 14, 2021}{Seoul, South Korea}
\acmBooktitle{9th Workshop on Encrypted Computing \& Applied Homomorphic Cryptography (WAHC'19), November 14, 2021, Seoul, South Korea}
\acmPrice{15.00}
\acmDOI{TODO}
\acmISBN{TODO}
\fi

\begin{document}
\renewcommand{\thelstlisting}{\arabic{lstlisting}}

\newcommand{\intelR}{Intel\textregistered} 
\newcommand{\intel}{Intel} 
\newcommand{\firstprojectname}{\intelR{} HEXL} 
\newcommand{\projectname}{\intel{} HEXL} 
\newcommand{\longprojectname}{{\intelR{} Homomorphic Encryption Acceleration Library (\firstprojectname{})}}

\newcommand{\repourl}{\href{https://github.com/intel/hexl}{https://github.com/intel/hexl}}

\newcommand{\LongTitle}{\projectname: Accelerating Homomorphic Encryption with Intel AVX512-IFMA52}
\newcommand{\ShortTitle}{\projectname}

\ifiacr
\title{\LongTitle}

\author{Anonymous for initial submissiom}
\institute{No institute for initial submission}


\maketitle

\else


\title[\ShortTitle]{\LongTitle}

\author{Fabian Boemer}
\email{fabian.boemer@intel.com}
\affiliation{%
\institution{Intel Corporation}
\city{Santa Clara}
\state{CA}
\country{USA}
}

\author{Sejun Kim}
\email{sejun.kim@intel.com}
\affiliation{%
\institution{Intel Corporation}
\city{Hillsboro}
\state{Oregon}
\country{USA}
}

\author{Gelila Seifu}
\email{gelila.seifu@intel.com}
\affiliation{%
\institution{Intel Corporation}
\city{Santa Clara}
\state{CA}
\country{USA}
}

\author{Fillipe D. M. de Souza}
\email{fillipe.souza@intel.com}
\affiliation{%
\institution{Intel Corporation}
\city{Folsom}
\state{CA}
\country{USA}
}

\author{Vinodh Gopal}
\email{vinodh.gopal@intel.com}
\affiliation{%
\institution{Intel Corporation}
\city{Hudson}
\state{MA}
\country{USA}
}

\fi

\newcommand\blfootnote[1]{%
	\begingroup
	\renewcommand\thefootnote{}\footnote{#1}%
	\addtocounter{footnote}{-1}%
	\endgroup
}

\begin{abstract}
\input{abstract}
\ifarxiv
 \blfootnote{\textcopyright Notice: Copyright \textcopyright 2021, Intel Corporation. All Rights Reserved. Intel TM Notice: Intel, the Intel logo, and other Intel marks are trademarks of Intel Corporation or its subsidiaries. Other names and brands may be claimed as the property of others. No product or component can be absolutely secure. Performance/Benchmarking Disclaimer: Performance varies by use, configuration and other factors. Learn more at www.intel.com/performanceindex.
 Intel technologies may require enabled hardware, software or service activation. Your results may vary.
 No product or component can be absolutely secure.
 Performance results are based on testing as of dates shown in configurations and may not reflect all publicly available updates.}
 \fi
 \ifiacr 
 \footnote{\textcopyright Notice: Copyright \textcopyright 2021, Intel Corporation. All Rights Reserved.
Intel TM Notice: Intel, the Intel logo, and other Intel marks are trademarks of Intel Corporation or its subsidiaries.
Other names and brands may be claimed as the property of others.
No product or component can be absolutely secure.
Performance/Benchmarking Disclaimer: Performance varies by use, configuration and other factors. Learn more at www.intel.com/performanceindex.
Intel technologies may require enabled hardware, software or service activation. Your results may vary.
No product or component can be absolutely secure.
Performance results are based on testing as of dates shown in configurations and may not reflect all publicly available updates.
}

 \fi
\end{abstract}


\ifiacr
\else
\fi

%
\ifiacr
\else
\maketitle
\fi




\input{body}


\clearpage
\ifiacr
\bibliographystyle{splncs04}
\else
\bibliographystyle{ACM-Reference-Format}

\fi
\bibliography{ref}



\end{document}

%% file: abstract.tex
Modern implementations of homomorphic encryption (HE) rely heavily on polynomial arithmetic over a finite field.
This is particularly true of the BGV, BFV, and CKKS HE schemes.
Two of the biggest performance bottlenecks in HE primitives and applications are polynomial modular multiplication and the forward and inverse number-theoretic transform (NTT).
Here, we introduce \emph{\longprojectname{}}, a C++ library which provides optimized implementations of polynomial arithmetic for \intelR{} processors.
\projectname{} takes advantage of the recent \intelR{} Advanced Vector Extensions 512 (\intelR{} AVX512) instruction set to provide state-of-the-art implementations of the NTT and modular multiplication.
On the forward and inverse NTT, \projectname{} provides up to 7.2x and 6.7x single-threaded speedup, respectively, over a native C++ implementation.
\projectname{} also provides up to 6.0x single-threaded speedup on the element-wise vector-vector modular multiplication, and 1.7x single-threaded speedup on the element-wise vector-scalar modular multiplication.
\projectname{} is available open-source at \repourl{} under the Apache 2.0 license and has been adopted by the Microsoft SEAL and PALISADE homomorphic encryption libraries.

%% file: body.tex
\section{Introduction}
Homomorphic encryption (HE) is a form of encryption which enables computation in the encrypted domain.
Homomorphic encryption is useful in applications with sensitive data, particularly medical and financial settings~\cite{bergamaschi2019homomorphic,kocabas2020towards,blatt2020secure}.
However, HE currently suffers from enormous memory and runtime overheads of up to 30,000x~\cite{jung2020heaan}.

Ciphertexts in many HE schemes, including BGV~\cite{brakerski2014leveled}, BFV~\cite{fan2012somewhat}, and CKKS~\cite{cheon2017homomorphic,cheon2018full}, are polynomials in finite fields, whose coefficients can be hundreds of bits and whose degree is typically a power in the range $[2^{10}, 2^{17}]$.
Performing HE computations requires operating on these large polynomials.
While recent years have seen tremendous improvement in HE performance due to algorithmic changes and optimized implementations, the performance overhead remains perhaps the biggest bottleneck to adoption of HE.

Here, we introduce \emph{\firstprojectname{}}, an open-source library which provides efficient implementations of integer arithmetic on finite fields.
\projectname{} targets polynomial operations with word-sized primes on 64-bit processors.
For efficient implementation, \projectname{} uses the \intel{} Advanced Vector Extensions 512 (\intel{} AVX512) instruction set to provide optimized implementations on \intel{} processors.
In particular, the \intelR{} Advanced Vector Extensions 512 Integer Fused Multiply Add (\intelR{} AVX512-IFMA52) instructions introduced in the 3rd Gen \intelR{} Xeon\textregistered\ Scalable Processors provide significant speedup on primes below 50\textendash{}52 bits.

We begin with a brief introduction of the mathematical concepts implemented in \projectname{} (Section~\ref{sec:Background}) and the Intel AVX512 instruction set (Section~\ref{sec:AVX512}).
Section~\ref{sec:previous_work} compares \projectname{} to existing work, explaining \projectname{}'s unique contribution lies in the application of the Intel AVX512-IFMA52 instruction set to word-size finite field arithmetic, such as the number-theoretic transform (NTT).
Next, we introduce the design (Section~\ref{sec:design}) and implementation (Section~\ref{sec:implementation}) of \projectname{}.
In particular, we provide detailed descriptions of the forward (Section~\ref{sec:implementation:ntt:fwd}) and inverse (Section~\ref{sec:implementation:ntt:inv}) NTT and polynomial kernels (Section~\ref{sec:implementation:poly}), including vector-vector modular multiplication (Section~\ref{sec:implementation:poly:eltwise_vector_vector_modmul}) and vector-scalar modular multiplication (Section~\ref{sec:implementation:poly:eltwise_vector_scalar_modmul}).
In Section~\ref{sec:he_library_integration}, we describe the integration of \projectname{} to two public HE libraries, Microsoft SEAL~\cite{sealcrypto} and PALISADE~\cite{palisade}.
We next demonstrate the performance of \projectname{} in Section~\ref{sec:results}, showcasing up to 7.2x and 6.7x speedup over a native C++ implementation of the forward and inverse NTT, respectively, up to 6.0x on element-wise vector-vector modular multiplication and 1.7x speedup on the element-wise vector-scalar modular multiplication.
Finally, we conclude in Section~\ref{sec:conclusion}.

\section{Background} \label{sec:Background}
We provide a brief background of the algorithms optimized used in \projectname{}, which are common building blocks of lattice cryptography.
Let $\mathcal{R}_q = \mathbb{Z}_q[X] / (X^N+1)$ be the polynomial quotient ring consisting of polynomials with degree at most $N-1$ and integer coefficients in the finite field $\mathbb{Z}_q = \{0, 1, \hdots, q-1\}$, where $q$ is a word-sized prime satisfying $q \equiv 1 \mod 2N$.
One way to represent a polynomial $a = \sum_{i=0}^{N-1} a_i x^i \in \mathcal{R}_q$ is via the coefficient embedding, i.e.
$$a = (a_0, a_1, a_2, \hdots, a_{N-1})$$
with $a_i \in \mathbb{Z}_q$.

Typical HE operations compute  on polynomials in $\mathcal{R}_q$ as follows:
\begin{itemize}
    \item \emph{Element-wise addition}. Given $a, b \in \mathcal{R}_q$, compute $c=a+b$ such that $c_i = (a_i + b_i) \mod q.$

    \item \emph{Element-wise negation}. Given $a \in \mathcal{R}_q$, compute $b = -a$ such that $b_i = q-a_i$.

    \item \emph{Element-wise multiplication}. Given $a, b \in \mathcal{R}_q$, compute $c=a \odot b$ such that $c_i = (a_i \cdot b_i) \mod q.$

    \item \emph{Element-wise vector-scalar multiplication}. Given $a \in \mathcal{R}_q, b \in \mathbb{Z}_q$, compute $c = a \cdot b$ such that $c_i = (a_i \cdot b) \mod q$.

    \item \emph{Vector-vector multiplication}. Given $a, b \in \mathcal{R}_q$, compute $c = a * b$ such that
    $$c_i = \sum_{j=0}^i a_j \cdot b_{i-j} - \sum_{j=i+1}^{N-1} a_j \cdot b_{N+i-j}.$$
    Note,
     $$X^N \equiv -1 \mod (X^N+1),$$ which yields the negative coefficients.
\end{itemize}
In practice, element-wise addition and element-wise negation are typically much faster to compute than the types of multiplication.
As such, we focus on the various multiplication functions in $\mathcal{R}_q$.

\subsection{Barrett Reduction}
Scalar modular multiplication is a primary bottleneck in lattice cryptography.
A simple implementation of scalar modular multiplication uses the 128-bit integer extension, supported by many modern compilers including gcc and clang.
The modulus operator \% is used for modular reduction.
Listing~\ref{lst:barrett} shows C/C++ source code for a simple implementation of scalar modular multiplication.
\begin{lstlisting}[language=C++, label=lst:barrett, caption={Native modular multiplication}]
// Returns (a * b) mod q
uint64_t naive_modmul(uint64_t a, uint64_t b, uint64_t q) {
    return uint64_t(uint128_t(a) * b) % q;
}
\end{lstlisting}
Performance of this naive implementation is poor due to the modulus operator, which will perform integer division via, e.g., the extended Euclidean algorithm.

In typical HE applications, the modulus $q$ is re-used for many modular multiplications.
In this setting, \emph{Barrett reduction} can be used to improve performance.
Barrett reduction takes advantage of the fact that
    $$x \mod q = x - \lfloor x / q \rfloor q$$
when $x/q$ is computed exactly.
If $x / q$ is computed with sufficient accuracy, the result remains correct.
Barrett reduction uses a pre-computed integer, $k$, based on the modulus $q$, to replace division with bit shifting.
This approximation requires an extra conditional subtraction to guarantee correctness.
Nevertheless, replacing integer division with bit shifting results in a speedup.
Algorithm~\ref{alg:barrett} shows Barrett's algorithm, as presented in~\cite{geraud2016double}.
\begin{algorithm}
    \caption{Barrett Reduction}
    \label{alg:barrett}
    \begin{algorithmic}[1]
        \Require{$q < 2^Q, d < 2^D, k = \big \lfloor \frac{2^L}{q} \big \rfloor$, with $Q \leq D \leq L$}
        \Ensure{Returns $d \mod q$}
        \Function{Barrett Reduction}{$d, q, k, Q, L$}
            \State{$c_1 \gets d \gg (Q - 1)$}
            \State{$c_2 \gets c_1 k$} \label{alg:barrett:c2}
            \State{$c_3 \gets c_2 \gg (L - Q + 1)$} \label{alg:barrett:c3}
            \State{$c_4 \gets d - q c_3$}
            \If{$c_4 \geq q$}
                \State{$c_4 \gets c_4 - q$}
            \EndIf
        \State \Return $c_4$
        \EndFunction
    \end{algorithmic}
\end{algorithm}

\subsection{Number-Theoretic Transform (NTT)}
\label{sec:background-ntt}
The number-theoretic transform (NTT) is another performance bottleneck in typical lattice cryptography computations.
The NTT is equivalent to the fast Fourier transform (FFT) in a finite field, i.e. all addition and multiplications are performed with respect to the modulus $q$.
Let $\omega$ be a primitive $N$'th root of unity in $\mathbb{Z}_q$ and $a = (a_0, \hdots, a_{N-1}) \in \mathbb{Z}^N_q$.
Then, the forward cyclic NTT is defined as $\tilde{a} = \text{NTT}(a)$, where $\tilde{a}_i = \sum_{j=0}^{N-1} a_j \omega^{ij} \mod q$ for $i=0, 1, \hdots, N-1$.
The inverse cyclic NTT is given by $b = InvNTT(\tilde{a})$, where $b_i = \frac{1}{n} \sum_{j=0}^{N-1} \tilde{a}_j \omega^{-ij} \mod q$ for $i=0, 1, \hdots, N-1$.
Note, $InvNTT(NTT(a)) = a$.

The NTT can be used to speed up polynomial-polynomial multiplication in $\mathcal{R}_q$.
However, using $\odot$ to indicate element-wise multiplication, the straightforward usage
$$InvNTT(NTT(a) \odot NTT(b))$$
corresponds to polynomial-polynomial multiplication in $\mathbb{Z}_q^N / (X^N - 1)$, whereas HE operates in $\mathcal{R}_q = \mathbb{Z}_q^N / (X^N + 1)$.
As described in~\cite{longa2016speeding}, a modification of the cyclic NTT, known as the \emph{negacyclic NTT}, or \emph{negative wrapped convolution}, can be used to perform polynomial multiplication in $\mathcal{R}_q$.
Let $\psi$ be a primitive $2N$'th root of unity in $\mathbb{Z}_q$.
Let $a, b \in \mathbb{Z}_q^N$ and $\hat{a} = (a_0, \psi a_1, \psi^2 a_2, \hdots, \psi^{N-1} a_{N-1})$, $\hat{b} = (b_0, \psi b_1, \psi^2 b_2, \hdots, \psi^{N-1} b_{n-1})$.
Then, the negacyclic NTT is defined as
$$c = (1, \psi^{-1}, \hdots, \psi^{-(N-1)}) \odot InvNTT(NTT(\hat{a}) \odot NTT(\hat{b}))),$$
 which satisfies $c = a * b$ in $\mathcal{R}_q$.
 The NTT-based formulation reduces the runtime of polynomial-polynomial modular multiplication from $O(N^2)$ to $O(N \log N)$.

 \ifiacr
 \else
\textbf{Optimized Implementation.}
\fi
The NTT inherits a rich history of optimizations from the FFT, in addition to several NTT-specific optimizations.
Similar to the FFT, the NTT has a recursive formulation attributed to Cooley and Tukey~\cite{cooley1965algorithm}.
Cooley-Tukey NTTs decompose an NTT of size $N = N_1 N_2$ as $N_1$ NTTs of size $N_2$ followed by $N_2$ NTTs of size $N_1$.
This recursive formulation reduces the runtime of the NTT to $O(N \log N)$, improving upon the $O(N^2)$ runtime of a naive implementation.
The choice of $\min(N_1, N_2)$ determines the \emph{radix} of the implementation.
One byproduct of the Cooley-Tukey forward NTT is that the output is in bit-reversed order.
That is, given an index $i$ in binary representation
$$i = 0bi_0i_1 \hdots i_{\log_2(N)} \in \{0,1\}^{\log_2(N)},$$
the output of the Cooley-Tukey NTT at index $i$ is
$$NTT(a)[0bi_{\log_2(N)} i_{\log_2(N)-1} \hdots i_1 i_0].$$
The inverse transform restores the standard bit ordering.
Typically, any operations performed in the bit-reversed domain are performed element-wise, so the bit-reversal usually does not pose a problem.
Furthermore, the Cooley-Tukey NTT may operate in-place, i.e. the output overwrites the input.
Algorithm~\ref{alg:cooley_tukey_fwd_ntt} shows a simple radix-2 in-place Cooley-Tukey NTT algorithm, taken from~\cite{longa2016speeding}.
Algorithm~\ref{alg:cooley_tukey_inv_ntt} shows the analogous radix-2 in-place Gentleman-Sande inverse NTT algorithm.
\begin{algorithm}
    \caption{Cooley-Tukey Radix-2 NTT}
    \label{alg:cooley_tukey_fwd_ntt}
    \begin{algorithmic}[1]
        \Require{$a = (a_0, a_1, \hdots, a_{N-1}) \in \mathbb{Z}_q^N$ in standard ordering. $N$ is a power of 2. $q$ is a prime satisfying $q \equiv 1 \mod 2N$. $\psi_\text{rev} \in \mathbb{Z}_q^N$ stores the powers of $\psi$ in bit-reversed order.}
        \Ensure{$a \gets NTT(a)$ in bit-reversed order.}
        \Function{Cooley-Tukey Radix-2 NTT}{$a, N, q, \psi_\text{rev}$}
            \State $t \gets n$
            \For{($m=1; m<n; m=2n$)}
                \State $t \gets t / 2$
                \For{($i=0; i < m; i$++)}
                    \State $j_1 \gets 2 \cdot i \cdot t$
                    \State $j_2 \gets j_1 + t - 1$
                    \State $W \gets \psi_{rev}[m+i]$
                    \For{$(j=j_1; j \leq j_2; j$++)} \label{alg:ct_ntt:loop_start}
                        \State $X_0 \gets a_j$ \label{alg:ct_ntt:radix_start}
                        \State $X_1 \gets a_{j+t}$
                        \State $a_j \gets X_0 + W \cdot X_1 \mod q$
                        \State $a_{j+t} \gets X_0 - W \cdot X_1 \mod q$ \label{alg:ct_ntt:radix_end}
                    \EndFor \label{alg:ct_ntt:loop_end}
                \EndFor
            \EndFor
            \State \Return $a$
        \EndFunction
    \end{algorithmic}
\end{algorithm}

\begin{algorithm}
    \caption{Gentleman-Sande (GS) Radix-2 InvNTT}
    \label{alg:cooley_tukey_inv_ntt}
    \begin{algorithmic}[1]
        \Require{$a = (a_0, a_1, \hdots, a_{N-1}) \in \mathbb{Z}_q^N$ in bit-reversed ordering. $N$ is a power of 2.$q$ is a prime satisfying $q \equiv 1 \mod 2N$. $\psi^{-1}_\text{rev} \in \mathbb{Z}_q^N$ stores the powers of $\psi^{-1}$ in bit-reversed order.}
        \Ensure{$a \gets InvNTT(a)$ in standard ordering.}
        \Function{Gentleman-Sande Radix-2 InvNTT}{$a, N, q, \psi_\text{rev}$}
            \State $t \gets 1$
            \For{($m=n; m > 1; m = m / 2$)}
                \State $j_1 \gets 0 $
                \State $h \gets m / 2$
                \For{($i=0; i < h; i$++)}
                    \State $j_2 \gets j_1 + t - 1$
                    \State $W \gets \psi_{rev}^{-1}[h+i]$
                    \For{$(j=j_1; j \leq j_2; j$++)} \label{alg:gs_ntt:loop_start}
                        \State $X_0 \gets a_j$ \label{alg:gs_intt:radix_start}
                        \State $X_1 \gets a_{j+t}$
                        \State $a_j \gets X_0 + X_1 \mod q$
                        \State $a_{j+t} \gets (X_0 - X_1) \cdot W \mod q$ \label{alg:gs_intt:radix_end}
                    \EndFor  \label{alg:gs_ntt:loop_end}
                    \State $j_1 \gets j_1 + 2t$
                \EndFor
                \State $t \gets 2t$
            \EndFor
            \For{($j = 0; j < n; j$++)}
                \State $a[j] \gets a[j] \cdot n^{-1} \mod q$
            \EndFor
            \State \Return $a$
        \EndFunction
    \end{algorithmic}
\end{algorithm}

The \emph{butterfly} refers to the radix-$r = \min(N_1, N_2)$ NTT.
For instance, the butterfly for the radix-2 NTT in Algorithm~\ref{alg:cooley_tukey_fwd_ntt} is given in lines~\ref{alg:ct_ntt:radix_start}\textendash{}\ref{alg:ct_ntt:radix_end}, which compute
$$(X_0, X_1) \mapsto (X_0 + WX_1 \mod q, X_0 - WX_1 \mod q).$$
Harvey~\cite{harvey2014faster} provides an optimization to the butterfly using a redundant representation $X_0,X_1 \in \mathbb{Z}_{4q}$.
Algorithm~\ref{alg:harvey_fwd_butterfly} shows the Harvey forward NTT butterfly\footnote{Note,~\cite{harvey2014faster} presents Algorithm~\ref{alg:cooley_tukey_fwd_ntt} as the inverse butterfly, whereas \projectname{} uses it for the forward NTT. This difference stems from the choice of `decimation-in-time' vs. `decimation-in-frequency.' The same applies for the inverse butterfly.}.
\begin{algorithm}
    \caption{Harvey NTT butterfly. $\beta$ is the word size, e.g. $\beta=2^{64}$ on typical modern CPU platforms.}
    \label{alg:harvey_fwd_butterfly}
    \begin{algorithmic}[1]
        \Require{$q < \beta/4$; $0 < W < q$}
        \Require{$W' = \lfloor W \beta / q \rfloor$, $0 < W' < \beta$}
        \Require{$0 \leq X_0, X_1 < 4q$}
        \Ensure{$Y_0 \gets X_0 + W X_1 \mod q; 0 \leq Y_0 < 4q$}
        \Ensure{$Y_1 \gets X_0 - W X_1 \mod q; 0 \leq Y_1 < 4q$}
        \Function{HarveyNTTButterfly}{$X_0, X_1, W, W', q, \beta$}
            \If{$X_0 \geq 2q$}
                \State $X_0 \gets X_0 - 2q$
            \EndIf
            \State $Q \gets \lfloor W' X_1 / \beta \rfloor$
            \State $T \gets (WX_1 - Qq) \mod \beta$
            \State $Y_0 \gets X_0 + T$
            \State $Y_1 \gets X_0 - T + 2q$
            \State \Return $Y_0, Y_1$
        \EndFunction
    \end{algorithmic}
\end{algorithm}
Using the Harvey butterfly in the Cooley-Tukey NTT yields outputs in $\mathbb{Z}^N_{4q}$, so an additional correction step is required to reduce the output to $\mathbb{Z}^N_q$.
Similar to the forward transform, Harvey~\cite{harvey2014faster} also provides an efficient butterfly for the inverse NTT, using a redundant representation in $[0, 2q)$.
Algorithm~\ref{alg:harvey_inv_butterfly} shows the Harvey inverse NTT butterfly.
\begin{algorithm}
    \caption{Harvey inverse NTT butterfly. $\beta$ is the word size, e.g. $\beta=2^{64}$ on typical modern CPU platforms.}
    \label{alg:harvey_inv_butterfly}
    \begin{algorithmic}[1]
        \Require{$q < \beta/4$; $0 < W < q$}
        \Require{$W' = \lfloor W \beta / q \rfloor$; $0 < W' < \beta$}
        \Require{$0 \leq X_0, X_1 < 2q$ }
        \Ensure{$Y_0 \gets X_0 + X_1 \mod q; 0 \leq Y_0 < 2q$.}
        \Ensure{$Y_1 \gets W(X_0 - X_1) \mod q; 0 \leq Y_1 < 2q$.}
        \Function{HarveyInvNTTButterfly}{$X_0, X_1, W, W', q, \beta$}
            \State $Y_0 \gets X_0 + X_1$
            \If{$Y_0 \geq 2q$}
                \State $Y_0 \gets Y_0 - 2q$
            \EndIf
            \State $T \gets X_0 - X_1 + 2q$
            \State $Q \gets \lfloor W' T / \beta \rfloor$
            \State $Y_1 \gets (WT - Qq) \mod \beta$
            \State \Return $Y_0, Y_1$
        \EndFunction
    \end{algorithmic}
\end{algorithm}

\subsection{\intel{} Advanced Vector Extensions} \label{sec:AVX512}
The \intelR{} Advanced Vector Extensions (\intelR{} AVX) is a set of single-instruction multiple data (SIMD) instructions for the x86 architecture.
\intel{} AVX instructions enable simultaneous computation on chunks of data larger than typical word-sized chunks.
For instance, the legacy \intelR{} Streaming SIMD Extensions (\intelR{} SSE) operates on 128-bit data chunks.
The AVX2 instruction set expanded the SIMD capability to 256-bit data chunks.
In recent years, the \intel{} AVX512 instruction set further expanded the SIMD capability to 512-bit data chunks.
Each SIMD instruction set can use the data chunks to represent multiple smaller-width inputs.
For instance, \intel{} AVX512 intrinsics use the \texttt{\_\_m512i} datatype, which represents a packed 512-bit integer, which may represent eight 64-bit integers.

To employ these SIMD instructions, users may call the desired assembly function.
Additionally, for easier use, \intel{} provides a set of C/C++-compatible intrinsics, which compile to the relevant assembly instruction.
To understand the naming of \intel{} intrinsics, `epi' refers to \emph{extended packed integer} and `epu' refers to \emph{extended packed unsigned integer} and the last number indicates the number of bits.

For instance, the \intelR{} AVX512 Doubleword and Quadword (\intelR{} AVX512-DQ) extension contains the following intrinsic:
\begin{itemize}
    \item \texttt{\_\_m512i \_mm512\_mullo\_epi64 (\_\_m512i a, \_\_m512i b)}. Given packed 64-bit integers in $a$ and $b$, return the low 64 bits of the 128-bit product $a \cdot b$.
\end{itemize}
However, there is no matching \texttt{\_mm512\_mulhi\_epi64} instruction.
Instead, it may be emulated with, e.g. four \intel{} AVX512 32-bit multiplies, five \intel{} AVX512 64-bit adds and five \intel{} AVX512 64-bit shift instructions~\cite{jung2020heaan}.

The \intel{} AVX512-IFMA52 extension to \intel{} AVX512~\cite{intelavx512ifma} consists of several operations useful in lattice cryptography.
In particular, \intel{} AVX512-IFMA52 introduces the following intrinsics:

\begin{itemize}
    \item \texttt{\_\_m512i \_mm512\_madd52lo\_epu64 (\_\_m512i a, \_\_m512i b, \_\_m512i c)}.
    Given packed unsigned 52-bit integers in each 64-bit element of $b$ and $c$, compute the 104-bit product $b \cdot c$.
    Add the low 52 bits of the product to the packed unsigned 64-bit integers in $a$ and return the result.
    \item \texttt{\_\_m512i \_mm512\_madd52hi\_epu64 (\_\_m512i a, \_\_m512i b, \_\_m512i c)}.
    Given packed unsigned 52-bit integers in each 64-bit element of $b$ and $c$, compute the 104-bit product $b \cdot c$.
    Add the high 52 bits of the product to the packed unsigned 64-bit integers in $a$ and return the result.
\end{itemize}
\projectname{} also utilizes one intrinsic from the \intel{} AVX512 Vector Bit Manipulation Version 2 (\intel{} AVX512-VBMI2) instruction set:
\begin{itemize}
    \item \texttt{\_\_m512i \_mm512\_shrdi\_epi64 (\_\_m512i a, \_\_m512i b, int imm8)}.
    Given packed 64-bit integers in $b$ and $c$, concatenate them to a 128-bit intermediate result.
    Shift the result right by \emph{imm8} bits and return the lower 64 bits of the result.
\end{itemize}

\section{Previous Work} \label{sec:previous_work}
The key differentiating factor between \projectname{} and previous work is the use of the \intel{} AVX512-IFMA52 instruction set to accelerate finite field arithmetic, in particular the number-theoretic transform.
Apart from this contribution, \projectname{} utilizes several existing algorithms from previous work.

The Mathemagix library~\cite{hoeven2016modular} provides \intel{} AVX-accelerated implementations of modular integer arithmetic using a SIMD programming model.
NFLlib~\cite{aguilar2016nfllib} provides similar acceleration of primitives common to the ring $\mathbb{Z}_q / (X^N + 1)$ using \intel{} SSE and \intel{} AVX2 instructions.
However, neither Mathemagix nor NFLlib consider \intel{} AVX512 implementations using the \intel{} AVX512-IFMA52 instruction set.

Previous work~\cite{gueron2016accelerating,edamatsu2019accelerating} using \intel{} AVX512-IFMA52 focuses on accelerating big integer multiplication without modular multiplication.
Drucker and Gueron~\cite{drucker2019fast} use \intel{} AVX512-IFMA52 to accelerate large integer modular squaring via Montgomery multiplication.
In contrast, our work uses Barrett reduction and focuses on word-sized modular multiplication.
Furthermore, to our knowledge, \projectname{} is the first work to accelerate the NTT using \intel{} AVX512-IFMA52.

\section{\projectname{}} \label{sec}

\subsubsection{Design} \label{sec:design}
\projectname{} is an open-source C++11 library available under the Apache 2.0 license.
\projectname{} focuses on the case where $q < \beta = 2^{64}$, as implemented on a 64-bit word-sized CPU platform.
This restriction of is typical of HE implementations on CPU.
SEAL~\cite{sealcrypto} for instance, bounds all coefficient moduli to 61 bits.
GPUs implementations of HE, (e.g.~\cite{morshed2020cpu,jung2020heaan}) however, often restrict $q < 2^{32}$ since 64-bit support for integers is often restricted or emulated, yielding lower performance.
As such, the \projectname{} API uses unsigned 64-bit integers input vector types.
Unlike other libraries, however, \projectname{} does not currently provide a BigNum type for multi-precision arithmetic, such as is found in NTL~\cite{shoup2001ntl} or NFLlib~\cite{aguilar2016nfllib}.

\projectname{} consists of a class for the NTT functionality in addition to several free functions implementing element-wise modular arithmetic on word-sized primes.
The NTT class performs pre-computation for the roots of unity and their pre-computed factors during initialization.
The element-wise functions perform any pre-computations outside the critical loop, rendering it unnecessary for the end user to perform any pre-computation.
\projectname{} is single-threaded and thread safe.
Listing~\ref{lst:ntt} shows the application programming interface (API) for the NTT.
\begin{lstlisting}[language=C++, basicstyle=\tiny, caption={\projectname{} NTT class API}, label=lst:ntt]
    class NTT {
      public:
        /// Initializes an empty NTT object
        NTT();

        /// Destructs the NTT object
        ~NTT();

        /// Initializes an NTT object with degree degree and modulus q.
        /// @param[in] degree a.k.a. N. Size of the NTT transform. Must be a power of 2
        /// @param[in] q Prime modulus. Must satisfy q == 1 mod 2N
        /// @brief Performs pre-computation necessary for forward and inverse
        /// transforms
        NTT(uint64_t degree, uint64_t q);

        /// @brief Initializes an NTT object with degree degree and modulus q
        /// @param[in] degree a.k.a. N. Size of the NTT transform. Must be a power of 2
        /// @param[in] q Prime modulus. Must satisfy q == 1 mod 2N
        /// @param[in] root_of_unity 2N'th root of unity in Z_q
        /// @details  Performs pre-computation necessary for forward and inverse
        /// transforms
        NTT(uint64_t degree, uint64_t q, uint64_t root_of_unity);

        /// @brief Compute forward NTT. Results are bit-reversed.
        /// @param[out] result Stores the result
        /// @param[in] operand Data on which to compute the NTT
        /// @param[in] input_mod_factor Assume input operand are in [0,
        /// input_mod_factor * q). Must be 1, 2 or 4.
        /// @param[in] output_mod_factor Returns output operand in [0,
        /// output_mod_factor * q). Must be 1 or 4.
        void ComputeForward(uint64_t* result, const uint64_t* operand,
                            uint64_t input_mod_factor, uint64_t output_mod_factor);

        /// Compute inverse NTT. Results are bit-reversed.
        /// @param[out] result Stores the result
        /// @param[in] operand Data on which to compute the NTT
        /// @param[in] input_mod_factor Assume input operand are in [0,
        /// input_mod_factor * q). Must be 1 or 2.
        /// @param[in] output_mod_factor Returns output operand in [0,
        /// output_mod_factor * q). Must be 1 or 2.
        void ComputeInverse(uint64_t* result, const uint64_t* operand,
                            uint64_t input_mod_factor, uint64_t output_mod_factor);
    }
\end{lstlisting}

Listing~\ref{lst:free_functions} shows the API for the element-wise operations.

\begin{lstlisting}[language=C++, basicstyle=\tiny, caption={\projectname{} free function API}, label=lst:free_functions]
/// @brief Adds two vectors element-wise with modular reduction
/// @param[out] result Stores result
/// @param[in] operand1 Vector of elements to add. Each element must be less
/// than the modulus
/// @param[in] operand2 Vector of elements to add. Each element must be less
/// than the modulus
/// @param[in] n Number of elements in each vector
/// @param[in] modulus Modulus with which to perform modular reduction. Must be
/// in the range [2, 2^{63} - 1].
/// @details Computes operand1[i] = (operand1[i] + operand2[i]) mod modulus
/// for i=0, ..., n-1.
void EltwiseAddMod(uint64_t* result, const uint64_t* operand1,
                   const uint64_t* operand2, uint64_t n, uint64_t modulus);

/// @brief Computes fused multiply-add (arg1 * arg2 + arg3) mod modulus element-wise, broadcasting scalars to vectors.
/// @param[out] result Stores the result
/// @param[in] arg1 Vector to multiply
/// @param[in] arg2 Scalar to multiply
/// @param[in] arg3 Vector to add. Will not add if arg3 == nullptr
/// @param[in] n Number of elements in each vector
/// @param[in] modulus Modulus with which to perform modular reduction. Must be
/// in the range [2, 2^{61} - 1]
/// @param[in] input_mod_factor Assumes input elements are in [0,
/// input_mod_factor * q). Must be 1, 2, 4, or 8.
void EltwiseFMAMod(uint64_t* result, const uint64_t* arg1, uint64_t arg2,
                   const uint64_t* arg3, uint64_t n, uint64_t modulus,
                   uint64_t input_mod_factor);

/// @brief Multiplies two vectors element-wise with modular reduction
/// @param[in] result Result of element-wise multiplication
/// @param[in] operand1 Vector of elements to multiply. Each element must be
/// less than the modulus.
/// @param[in] operand2 Vector of elements to multiply. Each element must be
/// less than the modulus.
/// @param[in] n Number of elements in each vector
/// @param[in] modulus Modulus with which to perform modular reduction
/// @param[in] input_mod_factor Assumes input elements are in [0,
/// input_mod_factor * q) Must be 1, 2 or 4.
/// @details Computes result[i] = (operand1[i] * operand2[i]) mod modulus for i=0, ..., n - 1
void EltwiseMultMod(uint64_t* result, const uint64_t* operand1,
                    const uint64_t* operand2, uint64_t n, uint64_t modulus,
                    uint64_t input_mod_factor);
\end{lstlisting}

Several of the functions have input arguments \texttt{input\_mod\_factor} or \texttt{output\_mod\_factor}.
These allows for optimized implementations via lazy reduction.
For instance, given two polynomials $f(x), g(x) \in \mathcal{R}_q$ represented using the coefficient embedding, we can compute $f * g$ by leaving the outputs of the forward NTT in the range $[0, 4q)$.
This improves performance over the simplest choice of \texttt{input\_mod\_factor = 1, output\_mod\_factor = 1}.

\begin{lstlisting}[language=C++, basicstyle=\tiny, caption={Use of \texttt{input\_mod\_factor} to optimize vector-vector modular multiplication}, label=lst:vector_vector_mult_mod]
/// Compute f(x) * g(x)
/// @param[in] N Size of input vectors
/// @param[in] q Modulus
VectorVectorMultMod(uint64_t* out, uint64_t* f, uint64_t* g, uint64_t N, uint64_t q) {
  NTT(N, p).ComputeForward(f, f, 1, 4);
  NTT(N, p).ComputeForward(g, g, 1, 4);
  EltwiseMultMod(out, f, g, N, q, 4);
  NTT(N, p).ComputeInverse(out, out, 1, 1);
}
\end{lstlisting}

\subsubsection{Implementation} \label{sec:implementation}
The primary functionality of \projectname{} is to provide optimized \intel{} AVX512-DQ and \intel{} AVX512-IFMA52 implementations for the forward and inverse NTT, element-wise vector-vector multiplication and element-wise vector-scalar multiplication.
The \intel{} AVX512-IFMA52 implementations are valid on prime moduli less than 50\textendash{}52 bits, while the \intel{} AVX512-DQ implementations allow moduli up to ${\sim}$62 bits, where the exact conditions depend also on the \texttt{input\_mod\_factor}.
\projectname{} also provides a reference native C++ implementation for each kernel, which has reduced performance but ensures \projectname{} is compatible with non-AVX512 processors.
The choice of implementation is determined at runtime based on the CPU feature availability.
While currently \projectname{} always prefers \intel{} AVX512-IFMA52 implementations over \intel{} AVX512-DQ implementations over native implementations, a future optimization may be to determine the best implementation dynamically via a small number of trials upon initializing the library.
This would ensure performance is always at least on par with the native C++ implementation, which may be efficient due to the compiler performing auto-vectorization.

\projectname{} uses several general optimizations in the implementation. Loops are unrolled either manually or using a pre-processor directive, with a manually-tuned unrolling factor. Within manually-unrolled loops, instructions are reordered where possible for best pipelining. Some loops are hand-coded for common input vector lengths, e.g. $N=8192, 16384$. As much as possible, cross-lane dependencies such as shuffles and permutations are avoided. Where applicable, memory is allocated to 64-byte boundaries to improve loads and stores. For best performance, the user input to \projectname{} functions should also be aligned to 64-byte boundaries.

\projectname{} uses several \intel{} AVX512 helper functions, which are inlined for best performance.
Several functions take a template argument, either 52 or 64, which is evaluated at compile time.
These template parameters enable a unified implementation between primes less than 50\textendash{}52 bits and primes larger than 52 bits, with no performance degradation.
The following \intel{} AVX512 kernels are used in \projectname{}, where we use \texttt{m512i} to refer to the \texttt{\_\_m512i} datatype:
\begin{itemize}
    \item \texttt{m512i \_mm512\_hexl\_mullo\_epi<k>(m512i x, m512i y)}. \\
    Multiplies packed unsigned $k$-bit integers in each 64-bit element of $x$ and $y$ to perform a $2k$-bit intermediate result.
    Returns the low $k$-bit unsigned integer from the intermediate result.
    The implementation with $k=64$ calls to \texttt{\_mm512\_mullo\_epi64}.
    The implementation with $k=52$ calls \texttt{\_mm512\_madd52lo\_epu64} with the accumulator set to zero.

    \item \texttt{m512i \_mm512\_hexl\_mullo\_add\_epi<k>}(\texttt{m512i x}, \texttt{m512i y}, \texttt{m512i z}). \\
    Multiplies packed unsigned $k$-bit integers in each 64-bit element of $y$ and $z$ to perform a $2k$-bit intermediate result.
    Returns the low $k$-bit unsigned integer from the intermediate result added to the low $k$ bits of $x$.
    The implementation with $k=64$ requires one call to \texttt{\_mm512\_mullo\_epi64} and one call to \texttt{\_mm512\_add\_epi64}.
    The implementation with $k=52$ requires a single call to \texttt{\_mm512\_madd52lo\_epu64}.

    \item \texttt{m512i \_mm512\_hexl\_mulhi\_epi<k>(m512i x, m512i y)}. \\
    Multiplies packed unsigned $k$-bit integers in each 64-bit element of $x$ and $y$ to perform a $2k$-bit intermediate result.
    Returns the high $k$-bit unsigned integer from the intermediate result.
    The implementation with $k=64$ requires two 32-bit shuffles, four 32-bit multiplies, three right shift, four 64-bit additions and one packed logical and operation.
    The implementation with $k=52$ calls \texttt{\_mm512\_madd52hi\_epu64} with the accumulator set to zero.

    \item \texttt{m512i \_mm512\_hexl\_small\_mod\_epi64<k>}(\texttt{m512i x}, \texttt{m512i q}, \texttt{m512i* q\_times\_2}, \texttt{mm512i* q\_times\_4}). \\
    Given packed unsigned 64-bit integers in $x$ and $q$, with each integer $0 \leq x_i < k \cdot q_i$, where $k \in \{1, 2, 4, 8\}$, returns $x \mod q$.
    The implementation for $k=2$ uses the fact that for unsigned integers $x < 2q$,
    $$x \mod q = \begin{cases} x - q & x \geq q \\ x \end{cases} = \min(x-q, x),$$
    which calls \texttt{\_mm512\_sub\_epi64} once and \texttt{\_mm512\_min\_epu64} once.
    For $k=4$ and $k=8$, $\log_2k$ repeated calls are made to both \texttt{\_mm512\_sub\_epi64} and \texttt{\_mm512\_min\_epu64} which utilize the \texttt{q\_times\_2} ($k=4,8$) and \texttt{q\_times\_4} ($k=8$)inputs, which are required not to be nullptr in these cases.
    For instance, Listing~\ref{lst:small_mod} shows the implementation when $k=8$.
    The three statements map the input from the range $[0,8q)$ to $[0,4q)$, then to $[0, 2q)$, and finally to $[0,q)$.
    \begin{lstlisting}[language=C++, label=lst:small_mod, caption={Small-input modular reduction}]
// Fast computation of x mod q for x < 8q
__m512i _mm512_hexl_small_mod_epu64<8>(__m512i x, __m512i q, __m512i* q_times_2, __m512i* q_times_4) {
  x = _mm512_min_epu64(x, _mm512_sub_epi64(x, *q_times_4));
  x = _mm512_min_epu64(x, _mm512_sub_epi64(x, *q_times_2));
  return _mm512_min_epu64(x, _mm512_sub_epi64(x, q));
}
    \end{lstlisting}

    \item \texttt{m512i \_mm512\_hexl\_cmpge\_epu64}(\texttt{m512i x},\texttt{m512i y}, \texttt{uint64\_t v)}. \\
    Given packed unsigned 64-bit integers in $x, y$, returns a packed 64-bit integer with value $v$ in each element for which $x > y$ and 0 otherwise.
    The implementation makes one call to \texttt{\_mm512\_maskz\_broadcastq\_epi64} and one call to \texttt{\_mm512\_cmpge\_epu64\_mask}.
\end{itemize}

\subsection{NTT} \label{sec:implementation:ntt}
\projectname{} provides optimized \intel{} AVX512 implementations of the negacyclic number-theoretic transform (NTT) with bit-reversed outputs.
At a high level, the implementation follows the radix-2 implementation from Cooley-Tukey and Gentleman-Sande, using the Harvey butterflies (see Section~\ref{sec:background-ntt}).
In each case, the butterfly is implemented across all 8 lanes of an \intel{} AVX512 input vector of 64-bit integers.

\subsubsection{Forward NTT} \label{sec:implementation:ntt:fwd}
The forward NTT is implemented using the Cooley-Tukey radix-2 transform in Algorithm~\ref{alg:cooley_tukey_fwd_ntt}.
The key acceleration using \intel{} AVX512 is the loop in lines~\ref{alg:ct_ntt:loop_start} to \ref{alg:ct_ntt:loop_end}.
Algorithm~\ref{alg:avx512_harvey_fwd_butterfly} shows the Harvey forward NTT butterfly implemented in \intel{} AVX512.
\begin{algorithm}
    \caption{\intel{} AVX512 Harvey NTT butterfly. $\beta$ is the word size, with either $\beta=2^{52}$ or $\beta=2^{64}$.}
    \label{alg:avx512_harvey_fwd_butterfly}
    \begin{algorithmic}[1]
        \footnotesize
        \Require{$q < \beta/4$; $0 < W < q$}
        \Require{$W' = \lfloor W \beta / q \rfloor$, $0 < W' < \beta$}
        \Require{$0 \leq X, Y < 4q$}
        \Require{twice\_modulus contains $2q$ in all 8 lanes}
        \Require{neg\_modulus contains $-q$ in all 8 lanes}
        \Ensure{$X \gets X + W Y \mod q; 0 \leq Y < 4q$}
        \Ensure{$Y \gets X - W Y \mod q; 0 \leq Y < 4q$}
        \Function{HarveyNTTButterfly<int BitShift, bool InputLessThanMod>}{\_\_m512i* X, \_\_m512i* Y, \_\_m512i W\_op, \_\_m512i W\_precon, \_\_m512i neg\_modulus, \_\_m512i twice\_modulus}
            \If{!InputLessThanMod}
                \StateD{*X = _mm512_hexl_small_mod_epu64(*X, twice_modulus);}
            \EndIf
           \StateD{__m512i Q = _mm512_hexl_mulhi_epi<BitShift>(W_precon, *Y);}
           \StateD{__m512i W_Y = _mm512_hexl_mullo_epi<BitShift>(W_op, *Y); }
           \StateD{__m512i T = _mm512_hexl_mullo_add_epi<BitShift>(W_Y, Q, neg_modulus); }
           \If{BitShift == 52} \label{alg:avx512_harvey_butterfly:if52_start}
                \StateD{T = _mm512_and_epi64(T, _mm512_set1_epi64((1UL << 52) - 1));}
           \EndIf  \label{alg:avx512_harvey_butterfly:if52_end}
           \StateD{__m512i twice_mod_minus_T = _mm512_sub_epi64(twice_modulus, T); }
           \StateD{ *Y = _mm512_add_epi64(*X, twice_mod_minus_T); }
           \StateD{ *X = _mm512_add_epi64(*X, T); }
        \EndFunction
    \end{algorithmic}
\end{algorithm}

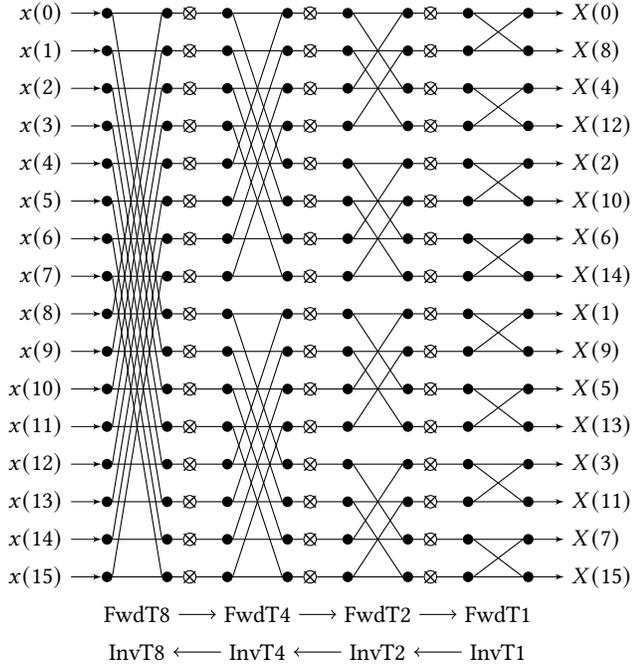
\begin{figure}
    \begin{center}
        \begin{tikzpicture}[yscale=0.5, xscale=0.8, node distance=0.3cm, auto]
            \tikzstyle{n}= [circle, fill, minimum size=4pt,inner sep=0pt, outer sep=0pt]
            \tikzstyle{mul} = [circle,draw,inner sep=-1pt]
            \newcounter{x}\newcounter{y}

            \foreach \y in {0,...,15}
                \node[n, pin={[pin edge={latex'-,black}]left:$x(\y)$}]
                    (N-0-\y) at (0,-\y) {};
            \foreach \y / \idx in {0/0,1/8,2/4,3/12,4/2,5/10,6,7/14,
                                8/1,9,10/5,11/13,12/3,13/11,14/7,15}
                \node[n, pin={[pin edge={-latex',black}]right:$X(\idx)$}]
                    (N-10-\y) at (7,-\y) {};
            \foreach \y in {0,...,15}
                \foreach \x / \c in {1/1,2/3,3/4,4/6,5/7,6/9}
                    \node[n, name=N-\c-\y] at (\x,-\y) {};
            \foreach \y in {0,...,15}
                \foreach \x / \c  in {1/2,4/5,7/8}
                    \node[mul, right of=N-\x-\y] (N-\c-\y) {${\times}$};

            \foreach \y in {0,...,15}
                \foreach \x in {0,1,3,4,6,7,9}
                {
                    \setcounter{x}{\x}\stepcounter{x}
                    \path (N-\x-\y) edge[-] (N-\arabic{x}-\y);
            }

            \setcounter{y}{0}
            \foreach \i / \j in {0/0,1/0,2/0,3/0,4/0,5/0,6/0,7/0,
                                    0/1,1/1,2/1,3/1,4/1,5/1,6/1,7/1}
            {
                \path (N-2-\arabic{y}) edge[-] node {} 
                        (N-3-\arabic{y});
                \stepcounter{y}
            }
            \setcounter{y}{0}
            \foreach \i / \j in {0/0,1/0,2/0,3/0,0/1,1/1,2/1,3/1,
                                0/0,1/0,2/0,3/0,0/1,1/1,2/1,3/1}
            {
                \path (N-5-\arabic{y}) edge[-] node {} 
                    (N-6-\arabic{y});
                \addtocounter{y}{1}
            }

            \setcounter{y}{0}
            \foreach \i / \j in {0/0,1/0,0/1,1/1,0/0,1/0,0/1,1/1,
                                    0/0,1/0,0/1,1/1,0/0,1/0,0/1,1/1}
            {
                \path (N-8-\arabic{y}) edge[-] node {} 
                    (N-9-\arabic{y});
                \stepcounter{y}
            }
            \foreach \sourcey / \desty in {0/8,1/9,2/10,3/11,
                                        4/12,5/13,6/14,7/15,
                                        8/0,9/1,10/2,11/3,
                                        12/4,13/5,14/6,15/7}
            \path (N-0-\sourcey.east) edge[-] (N-1-\desty.west);
            \foreach \sourcey / \desty in {0/4,1/5,2/6,3/7,
                                        4/0,5/1,6/2,7/3,
                                        8/12,9/13,10/14,11/15,
                                        12/8,13/9,14/10,15/11}
                \path (N-3-\sourcey.east) edge[-] (N-4-\desty.west);
            \foreach \sourcey / \desty in {0/2,1/3,2/0,3/1,
                                        4/6,5/7,6/4,7/5,
                                        8/10,9/11,10/8,11/9,
                                        12/14,13/15,14/12,15/13}
                \path (N-6-\sourcey.east) edge[-] (N-7-\desty.west);
            \foreach \sourcey / \desty in {0/1,1/0,2/3,3/2,
                                        4/5,5/4,6/7,7/6,
                                        8/9,9/8,10/11,11/10,
                                        12/13,13/12,14/15,15/14}
                \path (N-9-\sourcey.east) edge[-] (N-10-\desty.west);

            \node[name=FwdT8] at (1-0.5,-16) {FwdT8};
            \node[name=FwdT4] at (3-0.5,-16) {FwdT4};
            \node[name=FwdT2] at (5-0.5,-16) {FwdT2};
            \node[name=FwdT1] at (7-0.5,-16) {FwdT1};
            \draw[->] (FwdT8.east) -- (FwdT4.west);
            \draw[->] (FwdT4.east) -- (FwdT2.west);
            \draw[->] (FwdT2.east) -- (FwdT1.west);
            \node[name=InvT8] at (1-0.5,-17) {InvT8};
            \node[name=InvT4] at (3-0.5,-17) {InvT4};
            \node[name=InvT2] at (5-0.5,-17) {InvT2};
            \node[name=InvT1] at (7-0.5,-17) {InvT1};
            \draw[<-] (InvT8.east) -- (InvT4.west);
            \draw[<-] (InvT4.east) -- (InvT2.west);
            \draw[<-] (InvT2.east) -- (InvT1.west);
        \end{tikzpicture}

        \caption{NTT dataflow with root of unity factors omitted for clarity. The forward transform dataflow is from left to right. \emph{FwdT8} refers to a case when the \texttt{for} loop in Lines~\ref{alg:ct_ntt:loop_start}\textendash{}\ref{alg:ct_ntt:loop_end} of Algorithm~\ref{alg:cooley_tukey_fwd_ntt} runs for more than 8 iterations. Similarly, the \emph{FwdT4} corresponds to a loop with 4 iterations, \emph{FwdT2} corresponds to a loop with 2 iterations and \emph{FwdT1} corresponds to a loop with 1 iteration. The inverse forward transform dataflow is from right to left, with \emph{InvT8, InvT4, InvT2, InvT1} named analogously. Modified from~\cite{texamples}.}

        \label{fig:ntt_dataflow}
    \end{center}
\end{figure}

\subsubsection{Inverse NTT} \label{sec:implementation:ntt:inv}
The inverse NTT is implemented using the Gentleman-Sande radix-2 implementation from Algorithm~\ref{alg:cooley_tukey_inv_ntt}.
The key acceleration using \intel{} AVX512 is the loop in lines~\ref{alg:gs_ntt:loop_start} to \ref{alg:gs_ntt:loop_end}.
Algorithm~\ref{alg:avx512_harvey_inv_butterfly} shows the inverse Harvey NTT butterfly implemented in \intel{} AVX512.
\begin{algorithm}
    \caption{\intel{} AVX512 Harvey Inverse NTT butterfly. $\beta$ is the word size, with either $\beta=2^{52}$ or $\beta=2^{64}$.}
    \label{alg:avx512_harvey_inv_butterfly}
    \begin{algorithmic}[1]
        \footnotesize
        \Require{$q < \beta/4$; $0 < W < q$}
        \Require{$W' = \lfloor W \beta / q \rfloor$, $0 < W' < \beta$}
        \Require{ $0 \leq X, Y < 2q$ }
        \Require{twice\_modulus contains $2q$ in all 8 lanes}
        \Require{neg\_modulus contains $-q$ in all 8 lanes}
        \Ensure{$X \gets X + Y \mod q; 0 \leq Y < 2q$}
        \Ensure{$Y \gets W(X - Y) \mod q; 0 \leq Y < 2q$.}
        \Function{HarveyInvNTTButterfly<int BitShift, bool InputLessThanMod>}{\_\_m512i* X, \_\_m512i* Y, \_\_m512i W\_op, \_\_m512i W\_precon, \_\_m512i neg\_modulus, \_\_m512i twice\_modulus}
            \StateD{__m512i Y_minus_2q = _mm512_sub_epi64(*Y, twice_modulus);}
            \StateD{__m512i T = _mm512_sub_epi64(*X, Y_minus_2q);}

            \If{InputLessThanMod}
                \StateD{*X = _mm512_add_epi64(*X, *Y);}
            \Else
                \StateD{ *X = _mm512_add_epi64(*X, Y_minus_2q); }
                \StateD{ __mmask8 sign_bits = _mm512_movepi64_mask(*X); }
                \StateD{ *X = _mm512_mask_add_epi64(*X, sign_bits, *X, twice_modulus); }
            \EndIf
           \StateD{__m512i Q = _mm512_hexl_mulhi_epi<BitShift>(W_precon, T);}
           \StateD{__m512i Q_p = _mm512_hexl_mullo_epi<BitShift>(Q, neg_modulus); }
           \StateD{*Y = _mm512_hexl_mullo_add_epi<BitShift>(Q_p, W_op, T); }
           \If{BitShift == 52} \label{alg:avx512_harvey_inv_butterfly:if52_start}
                \StateD{T = _mm512_and_epi64(T, _mm512_set1_epi64((1UL << 52) - 1));}
           \EndIf  \label{alg:avx512_harvey_inv_butterfly:if52_end}
        \EndFunction
    \end{algorithmic}
\end{algorithm}
We make a few remarks on the implementations:
\begin{itemize}
    \item The template argument \texttt{InputLessThanMod} enables a compile-time optimization in when the inputs $X, Y$ are known to be less than $q$.
    For instance, when the input polynomial to the forward or inverse NTT has all coefficients less than $q$, \texttt{InputLessThanMod} is true during the first pass through the data.

    \item The negated modulus $-q$ is passed to the input of the forward and inverse butterflies.
    This enables the use of \texttt{\_mm512\_hexl\_mullo\_add\_epi}, which is a single instruction when \texttt{BitShift} is 52.
    Note, the \intel{} AVX512-IFMA52 instruction set does not contain an \texttt{\_mm512\_msub52lo\_epu64} instruction, which would enable using $q$ instead of $-q$.

    \item Lines~\ref{alg:avx512_harvey_butterfly:if52_start}\textendash{}\ref{alg:avx512_harvey_butterfly:if52_end} in the forward butterfly and Lines~\ref{alg:avx512_harvey_inv_butterfly:if52_start}\textendash{}\ref{alg:avx512_harvey_inv_butterfly:if52_end} in the inverse butterfly clear the high 12 bits from \texttt{T}.
    This is required because the \texttt{\_mm512\_madd52lo\_epu64} instruction uses a 64-bit accumulator, whereas our algorithm requires a 52-bit accumulator.

    \item In the inverse butterfly, rather than computing $Y_0 = X_0 + X_1; T = X_0 - X_1 + 2q$ (as presented in Algorithm~\ref{alg:harvey_inv_butterfly} and~\cite{harvey2014faster}), we compute \texttt{Y\_minus\_2q = Y - 2q; T = X - Y\_minus\_2q}. This saves two scalar additions at the cost of one extra scalar subtraction.
\end{itemize}

The \intel{} AVX512 NTT butterflies are used in every stage of the NTT.
The forward NTT \texttt{for} loop in Lines~\ref{alg:ct_ntt:loop_start} -~\ref{alg:ct_ntt:loop_end} of Algorithm~\ref{alg:cooley_tukey_fwd_ntt} begins with $N/2$ iterations in the first stage, then $N/4$ iterations in the next stage, followed by successive divisions by 2 until the final loop runs for a single iteration in the final stage.
As such, when the loop runs for 8 or more iterations, the use of Algorithm~\ref{alg:avx512_harvey_fwd_butterfly} is simple to apply (particularly since the number of loop iterations is divisible by 8).
However, special consideration must be taken in the final three stages, denoted \emph{FwdT4, FwdT2, FwdT1} in Figure~\ref{fig:ntt_dataflow} when the loop runs for 4 iterations, 2 iterations, and 1 iteration, respectively.
In these cases, we permute the data within each \intel{} AVX512 unit such that the butterfly is still applied across all lanes.

Similarly, the inverse NTT \texttt{for} loop in Lines~\ref{alg:gs_ntt:loop_start}\textendash{}\ref{alg:gs_ntt:loop_end} of Algorithm~\ref{alg:cooley_tukey_inv_ntt} begins with 1 iteration in the first stage, 2 iterations in the next stage, followed by successive multiplications by 2 until the final loop runs for $N/2$ iterations in the final stage.
Analogous to the forward NTT, for the first three stages, denoted \emph{InvT1, InvT2, InvT4} in Figure~\ref{fig:ntt_dataflow}, we permute the data within each \intel{} AVX512 unit such that the butterfly is still applied across all lanes.
We note the primary benefit of \intel{} AVX512-IFMA52 is in the NTT on primes less than 50 bits, in which case \texttt{\_mm512\_hexl\_mulhi<52>} via a single assembly call, whereas large primes require \texttt{\_mm512\_hexl\_mulhi<64>}, which is much more expensive (see Section~\ref{sec:AVX512}).

\subsection{Polynomial Kernels} \label{sec:implementation:poly}
For simplicity, in our presentation, we assume the degree of the input polynomials is divisible by 8.
This is typically the case for our HE applications, in which the polynomials are of degree $N = 2^k \geq 1024$ a power of two.
Nevertheless, the \projectname{} implementation includes logic that processes the $N \mod 8$ remaining loop iterations.

\subsubsection{Element-wise Vector-Vector Multiplication} \label{sec:implementation:poly:eltwise_vector_vector_modmul}
\projectname{} provides two AVX512 implementations of element-wise vector-vector multiplication: 1) an \intel{} AVX512-DQ implementation using integer logic; 2) an \intel{} AVX512-DQ implementation using floating-point logic.
For each implementation, we use a pre-processor directive to tune the loop unrolling factor for best performance.
Each \intel{} AVX512 kernel implements SIMD modular multiplication across all 8 lanes of the \intel{} AVX512 data and is sequentially applied to each 512-bit chunk of the input data.

\ifiacr
\paragraph{\intel{} AVX512-DQ integer implementation}
\else
\textbf{\intel{} AVX512-DQ integer implementation} \\
\fi
The \intel{} AVX512-DQ integer implementation uses Algorithm~\ref{alg:barrett}.
We choose $Q = \lfloor \log_2 q \rfloor + 1$ and $L = 63 + Q$.
This choice has a few benefits.
Firstly, this ensures $q > 2^{Q-1}$, which implies the Barrett factor $k = \lfloor 2^L / q \rfloor < 2^L / 2^{Q-1} = 2^{64}$, i.e. it fits in a single 64-bit integer.
Secondly, this choice ensures $L - Q + 1 = 64$, which implies $c_3$ is simply the high 64 bits of $c_2$ (see lines~\ref{alg:barrett:c2}\textendash{}\ref{alg:barrett:c3}), i.e. the low 64 bits of $c_2$ do not need to be computed.

Note, since the input may be larger than $q$ (when the \\\texttt{input\_mod\_factor} is larger than 1), the shifted product $d \gg (Q-1)$ may be larger than $2^{64}$.
To prevent overflow in this case (which may happen when $q$ exceeds 59 bits), the inputs are first reduced to the range $[0,q)$ via a sequence of fixed-time conditional subtractions.
Algorithm~\ref{alg:vector_vector_mod_mul_avx512_int} shows the pseudocode for the \intel{} AVX512-DQ integer modular multiplication implementation.
\begin{algorithm}
    \caption{VectorVectorModMulAVX512Int}
    \label{alg:vector_vector_mod_mul_avx512_int}
    \begin{algorithmic}[1]
        \footnotesize
        \Require{$q < 2^{62}$ stores the modulus in all 8 lanes}
        \Require{  $0 < X, Y < \texttt{InputModFactor} \cdot q$ }
        \Require{$\text{InputModFactor} \cdot q < 2^{63}$ }
        \Require{ \texttt{twice\_q} stores $2q$ across all 8 lanes }
        \Require{ \texttt{barr\_lo} stores $\lfloor {2^L} / q \rfloor$ across all 8 lanes}
        \Ensure{Returns $X \cdot Y \mod q$}
        \Function{EltwiseMultModAVX512Int<int BitShift, int InputModFactor>}{\_\_m512i X, \_\_m512i Y, \_\_m512i barr\_lo,
        \_\_m512i q, \_\_m512i twice\_q}
            \StateD{ X = _mm512_hexl_small_mod_epu64<InputModFactor>(X, q, twice_q); }
            \StateD{ Y = _mm512_hexl_small_mod_epu64<InputModFactor>(Y, q, twice_q); }
            \StateD{__m512i prod_hi = _mm512_hexl_mulhi_epi<64>(X, Y);}
            \StateD{__m512i prod_lo = _mm512_hexl_mullo_epi<64>(X, Y);}
            \StateD{__m512i c1 = _mm512_hexl_shrdi_epi64<BitShift - 1>(prod_lo, prod_hi);}
            \StateD{__m512i c3 = _mm512_hexl_mulhi_epi<64>(c1, barr_lo);}
            \StateD{__m512i c4 = _mm512_hexl_mullo_epi<64>(c3, q); }
            \StateD{c4 = _mm512_sub_epi64(prod_lo, c4); }
            \StateD{__m512i result = _mm512_hexl_small_mod_epu64(c4, q);}

            \State \Return result;
        \EndFunction
    \end{algorithmic}
\end{algorithm}

\ifiacr
\paragraph{\intel{} AVX512-DQ Floating-point Implementation}
\else
\textbf{\intel{} AVX512-DQ Floating-point Implementation} \\
\fi
For $q < 2^{50}$, \projectname{} uses a floating-point \intel{} AVX512-DQ implementation.
This implementation adapts Function 3.10 from Mathemagix~\cite{hoeven2016modular} to \intel{} AVX512, in a similar manner as Fortin et al.~\cite{fortin2020high}.
Algorithm~\ref{alg:vector_vector_mod_mul_avx512_float} shows the implementation for the \intel{} AVX512 floating-point kernel.
\begin{algorithm}
    \caption{VectorVectorModMulAVX512Float}
    \label{alg:vector_vector_mod_mul_avx512_float}
    \begin{algorithmic}[1]
    \footnotesize
        \Require{$0 < X, Y < \text{InputModFactor} \cdot q$}
        \Require{$\text{InputModFactor} \cdot q < 2^{52}$}
        \Require{ $u$ stores $1 / (\text{double})q$  in each lane, rounding toward infinity, so that $u \geq 1/q$}
        \Ensure{Returns $X \cdot Y \mod q$ }
        \Function{EltwiseMultModAVX512Float<int BitShift, int InputModFactor>}{\_\_m512i X, \_\_m512i Y, \_\_m512d u}
            \StateD{   const int rounding = _MM_FROUND_TO_POS_INF|_MM_FROUND_NO_EXC;}
            \StateD{   __m512d xi = _mm512_cvt_roundepu64_pd( X, rounding); }
            \StateD{   __m512d yi = _mm512_cvt_roundepu64_pd( Y, rounding); }
            \StateD{   __m512d h = _mm512_mul_pd(xi, yi); }
            \StateD{   __m512d l = _mm512_fmsub_pd(xi, yi, h); } \Comment{rounding error; h + l == x * y}
            \StateD{   __m512d b = _mm512_mul_pd(h, u); }      \Comment{ ${\sim}$(x * y) / q }
            \StateD{   __m512d c = _mm512_floor_pd(b);   }    \Comment{ ${\sim}$floor(x * y / q) }
            \StateD{   __m512d d = _mm512_fnmadd_pd(c, q, h); }
            \StateD{   __m512d g = _mm512_add_pd(d, l); }
            \StateD{   __mmask8 m = _mm512_cmp_pd_mask(g, _mm512_setzero_pd(), _CMP_LT_OQ); }
            \StateD{   g = _mm512_mask_add_pd(g, m, g, p); }
            \StateD{   __m512i result = _mm512_cvt_roundpd_epu64(g, rounding); }

            \State \Return result
        \EndFunction
    \end{algorithmic}
\end{algorithm}

We make a few notes about the floating-point implementation:
\begin{itemize}
    \item The implementation is valid as long as $\text{InputModFactor} \cdot q < 2^{52}$.
    As such, there is no explicit modulus reduction step required for $\text{InputModFactor} \in \{1, 2, 4\}$.
    \item We experimented with several \intel{} AVX512-IFMA52 implementations.
    However, we found this \intel{} AVX512 floating-point implementation yields best performance.
\end{itemize}

\subsubsection{Element-wise Vector-Scalar Multiplication} \label{sec:implementation:poly:eltwise_vector_scalar_modmul}
The \texttt{EltwiseFMAMod} function implements vector-scalar modular multiplication, with an additional optional scalar modular addition.
The \intel{} AVX512-DQ and \intel{} AVX512-IFMA52 implementations use the same underlying kernel, with the \intel{} AX512-DQ kernel using \texttt{BitShift = 64} and the \intel{} AVX512-IFMA52 kernel using \texttt{BitShift = 52}.
The \intel{} AVX512-IFMA52 kernel is valid for $\text{InputModFactor} \cdot q < 2^{52}$, while the \intel{} AVX512-DQ kernel is valid for $\texttt{InputModFactor} \cdot q < 2^{62}$.
Algorithm~\ref{alg:avx512_eltwise_fma_mod} shows the Algorithm for vector-scalar multiplication.
Compared to the vector-vector multiplication algorithm, the vector-scalar multiplication algorithm performs additional pre-computation using the scalar factor.
Where required, Lines \ref{alg:avx512_eltwise_fma_mod:reduce1} and \ref{alg:avx512_eltwise_fma_mod:reduce2} perform conditional subtractions to reduce the input to the range $[0, q)$.

\begin{algorithm}
    \caption{EltwiseFMAModAVX512}
    \label{alg:avx512_eltwise_fma_mod}
    \begin{algorithmic}[1]
        \footnotesize
        \Require{$0 < X, Z < \text{InputModFactor} \cdot q$}
        \Require{$0 < Y < q$}
        \Require{$Y_\text{barr} = \lfloor y \ll \text{BitShift} / q \rfloor$ }
        \Require{$\text{InputModFactor} \cdot q < 2^{\text{BitShift}}$}
        \Require{$q$ stores the modulus across all 8 lanes}
        \Ensure{Returns $X \cdot Y + Z \mod q$ }
        \Function{EltwiseFMAModAVX512<int BitShift, int InputModFactor>}{\_\_m512i X, \_\_m512i Y, \_\_m512i Y\_barr, \_\_m512i Z, \_\_m512i q}
            \StateD{ X = _mm512_hexl_small_mod_epu64<InputModFactor>(X, q); } \label{alg:avx512_eltwise_fma_mod:reduce1}
            \StateD{ Z = _mm512_hexl_small_mod_epu64<InputModFactor>(Z, q); } \label{alg:avx512_eltwise_fma_mod:reduce2}

            \StateD{ __m512i XY = _mm512_hexl_mullo_epi<64>(X, Y); }

            \StateD{ __m512i R = _mm512_hexl_mulhi_epi<BitShift>(X, Y_barr); }
            \StateD{ __m512i Rq = _mm512_mullo_epi64(R, q); }
            \StateD{ R = _mm512_sub_epi64(XY, Rq); }
            \StateD{ R = _mm512_hexl_small_mod_epu64(R, q); } \Comment{Conditional Barrett subtraction}

            \StateD{ R = _mm512_add_epi64(vq, Z); }
            \StateD{ R = _mm512_hexl_small_mod_epu64(R, q); }

            \State \Return R
        \EndFunction
    \end{algorithmic}
\end{algorithm}

\section{Integration with HE Libraries} \label{sec:he_library_integration}
Existing homomorphic encryption libraries such as Microsoft SEAL~\cite{sealcrypto} and PALISADE~\cite{palisade} typically provide a public API at the level of the HE scheme implemented, e.g. encryption, homomorphic multiplication and addition, and decryption.
This HE cryptography layer is typically implemented by calls to a lower-level polynomial arithmetic layer.
This polynomial arithmetic layer may implement $\mathcal{R}_q$ for a multi-word \emph{coefficient modulus} $q \in \mathbb{Z}_q$.
A common optimization is to represent a multi-word integer $x \mod q$ as a vector of pairwise coprime word-sized moduli $x \mod q \mapsto (x \mod q_1, x \mod q_2, \hdots, x \mod q_L)$.
This form is known as the residue number system (RNS) and has correctness guaranteed by the Chinese remainder theorem.
The benefit of the RNS form is that two numbers in RNS form can be multiplied in $\mathbb{Z}_q$ using the element-wise residues, yielding substantial speedup over a multi-word multiplication algorithm.

\projectname{} is designed to intercept HE libraries at the polynomial layer, with polynomials in RNS form.
We chose this as the target integration layer for a few reasons.
Firstly, the majority of the runtime in each HE operation lies at or below this polynomial layer, with minimal overhead from the HE layer to the polynomial layer.
Therefore, any speedup at the polynomial level will propagate to higher-level HE operations.
Secondly, the polynomial layer is usually common to different HE schemes, allowing \projectname{} to accelerate multiple HE schemes with a minimal integration surface to the HE library.
Figure~\ref{fig:he_integration} shows a high-level visualization of where \projectname{} integrates to the different layers in a standard HE library.

One downside to this layer of integration is the runtime is typically too small to effectively offload the computation to an accelerator such as a field-programmable gate array (FPGA) or graphic processing unit (GPU).
Another downside is that the optimization surface within \projectname{} is relatively small.
As such, for best performance, the HE library must still take care to optimize the sequence of calls to the polynomial layer, taking particular care to avoid many pitfalls in HE implementations, such as unnecessary NTT conversions and modular reductions, and cache misses when computing on multiple large polynomials.
Additionally, the HE library should align any input data to 64-byte boundaries aligned for best performance.

\projectname{} is publicly integrated with Microsoft SEAL~\cite{sealcrypto} since version 3.6.4 and PALISADE~\cite{palisade} since version v1.11.3.

\usetikzlibrary{arrows,positioning,shapes.geometric}
\begin{figure}
    \begin{center}
        \begin{tikzpicture}[>=latex']
            \tikzset{
                rblock/.style={draw, shape=rectangle,rounded corners=1.5em,align=center,minimum width=3cm,minimum height=1cm},
                topblock/.style={draw, shape=rectangle,rounded corners=1.5em,align=center,minimum width=8.5cm,minimum height=1cm},
                test/.style={draw, shape=diamond, aspect=2,}
            }
            \begin{scope}
                \node[topblock] (crypto_layer) {Cryptography layer - HE operations \\ {\small SEAL / PALISADE}};
                \node[test, below =0.5cm of crypto_layer] (hexl_config) {\projectname{} enabled?};
                \node[left=0.5cm of hexl_config] (hexl_config_no) {No};
                \node[right=0.5cm of hexl_config] (hexl_config_yes) {Yes};
                \node[rblock, below =0.5cm of hexl_config_no] (poly_layer) {Polynomial layer \\  {\small SEAL / PALISADE}};
                \node[rblock, below =0.5cm of hexl_config_yes] (hexl_layer) {Polynomial layer \\ {\small \projectname{}} };
                \node[rblock, below =0.5cm of poly_layer] (scalar_layer) {Scalar arithmetic \\  {\small SEAL / PALISADE}};
                \node[rblock, below =0.5cm of hexl_layer] (avx_layer) {AVX512 arithmetic \\ {\small \projectname{}} };
            \end{scope}

            \path[draw,->] (crypto_layer) edge (hexl_config);
            \path[draw,->] (hexl_config.west) -- (hexl_config_no) -- (poly_layer);
            \path[draw,->] (hexl_config.east) -- (hexl_config_yes) -- (hexl_layer);
            \path[draw,->] (poly_layer) edge (scalar_layer);
            \path[draw,->] (hexl_layer) edge (avx_layer);
        \end{tikzpicture}

    \caption{Architecture diagram of typical HE libraries showing \projectname{} integrates at the polynomial layer.}
    \label{fig:he_integration}
    \end{center}
\end{figure}
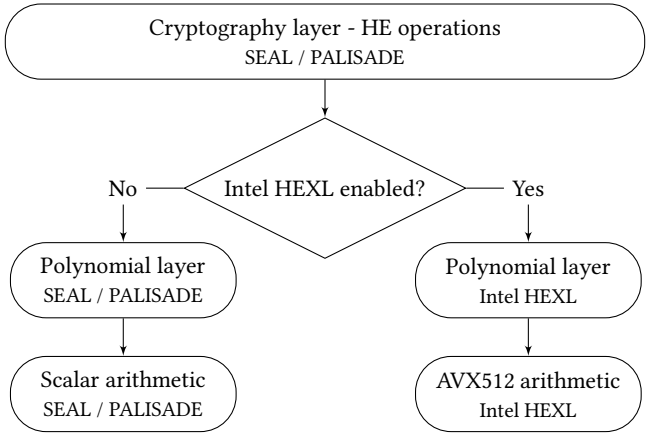

\section{Results} \label{sec:results}
We provide two levels of benchmarking.
First, we benchmark the low-level kernel implemented in \projectname{}.
We also integrate \projectname{} with two popular open-source libraries: Microsoft SEAL~\cite{sealcrypto} and PALISADE~\cite{palisade} and benchmark the higher-level HE operations within each library.
We benchmark each kernel on a 3rd Gen \intel{} Xeon\textregistered\ Scalable Processors Platinum 8360Y 2.4GHz processor with 64GB of RAM and 72 cores, running the Ubuntu 20.04 operating system.
The code is compiled using the clang-10 compiler with the `-march=native -O3' optimization flags.
Each benchmark runs single-threaded on a single core.

\subsection{\projectname{} Kernels}
\subsubsection{NTT}
We benchmark the performance of the  \projectname{} forward and inverse NTT in three settings: 1) the native C++ implementation; 2) the \intel{} AVX512-DQ implementation; 3) the \intel{} AVX512-IFMA52 implementation.
The default implementation uses the radix-2 Cooley-Tukey (forward NTT) and Gentleman-Sande (inverse NTT) formulations, using the Harvey butterfly (see Section~\ref{sec:background-ntt}).
Additionally, we compare against NTL v11.4.3~\cite{shoup2001ntl} and NFLlib~\cite{aguilar2016nfllib}, two open-source libraries implementing the NTT.
NTL is compiled with clang-10 using the \emph{NTL\_ENABLE\_AVX\_FFT} flag, which enables an experimental \intel{} AVX512 implementation using floating-point arithmetic. NFLlib is compiled with clang-10 using the \emph{NFL\_OPTIMIZED=ON} flag, which enables an \intel{} AVX256 implementation using 32-bit limbs.

We measure the performance on three different input sizes: $N=1024, 4096, 16384$.
Table~\ref{tab:results:fwdntt} shows the runtimes for the forward transform.
The \intel{} AVX512-DQ implementation provides a significant 2.7x\textendash{}2.8x speedup over the native implementation, with the \intel{} AVX512-IFMA52 increasing this speedup to 7.2x for the smallest size.
The AVX512-IFMA52 implementation speedup decreases to 5.3x for the larger $N=16384$ case as L1 cache misses bottleneck the memory access.
NFLlib's use of AVX256 yields performance between that of the native C++ and AVX512 implementations.
NTL's implementation uses floating-point arithmetic for integer computation, and is therefore correct only for primes up to 50 bits.
As such, while NTL may provide best performance on systems without \intel{} AVX512-IFMA52, no additional speedup is expected on NTL on systems with the \intel{} AVX512-IFMA52 instruction set.

\begin{table}
	\begin{center}
        \ifiacr
        \else
		\small
		\renewcommand\theadfont{\small}
        \fi
		\caption{Single-threaded, single-core runtime in microseconds of the forward NTT on a 50-bit modulus with \texttt{input\_mod\_factor} = \texttt{output\_mod\_factor} = 1.}
		\label{tab:results:fwdntt}
		\begin{tabular}{l l l l l l l}
            \toprule
            \multirow{2}{*}{Implementation} &
			\multicolumn{6}{c}{{$N$ / Speedup}} \\
            \cmidrule(lr){2-7}
                                            & 1024 &        & 4096 &         & 16384 &     \tabularnewline
            \midrule
            Native C++                      & 9.08 & 1.0x   & 38.8  & 1.0x   & 177  & 1.0x \tabularnewline
            NFLlib\cite{aguilar2016nfllib}  & 4.82 & 1.8x   & 21.1  & 1.8x   & 97.8 & 1.8x \tabularnewline
            \intel{} AVX512-DQ              & 3.26 & 2.7x   & 13.4  & 2.8x   & 62.3 & 2.8x \tabularnewline
            NTL\cite{shoup2001ntl}          & 2.44 & 3.7x   & 8.48  & 4.5x   & 40.2 & 4.3x \tabularnewline
            \intel{} AVX512-IFMA52          & 1.25 & 7.2x   & 5.81  & 6.6x   & 33.1 & 5.3x \tabularnewline
			\bottomrule
		\end{tabular}
	\end{center}
\end{table}

Table~\ref{tab:results:invntt} shows the runtimes for the inverse NTT.
We see the \intel{} AVX512-DQ implementation provides a similar speedup of 2.5x\textendash{}2.6x over the native implementation.
The \intel{} AVX512-IFMA52 implementation improves this speedup to 6.7x on the smaller transforms, which diminishes to 5.3x on the largest transform.
As with the forward NTT, the speedup on the larger transforms is diminished due to L1 cache misses.
Similarly, NFLlib's use of AVX256 yields performance between that of the native C++ and AVX512 implementations.
As with the forward transform, while NTL may provide best performance on systems without \intel{} AVX512-IFMA52, no additional speedup is expected on NTL on systems with the \intel{} AVX512-IFMA52 instruction set.

\begin{table}
	\begin{center}
        \ifiacr
        \else
		\small
		\renewcommand\theadfont{\small}
        \fi
		\caption{Single-threaded, single-core runtime in microseconds of the inverse NTT on a 50-bit modulus with \texttt{input\_mod\_factor} = \texttt{output\_mod\_factor} = 1.}
		\label{tab:results:invntt}
		\begin{tabular}{l l l l l l l}
            \toprule
            \multirow{2}{*}{Implementation} &
			\multicolumn{6}{c}{{$N$ / Speedup}} \\
            \cmidrule(lr){2-7}
                                            & 1024 &         & 4096 &         & 16384 &      \tabularnewline
            \midrule
            Native C++                      & 8.25 & 1.0x    & 37.8 & 1.0x    & 174   & 1.0x \tabularnewline
            NFLlib\cite{aguilar2016nfllib}  & 6.07 & 1.3x    & 26.7 & 1.4x    & 124   & 1.4x \tabularnewline
            \intel{} AVX512-DQ              & 3.16 & 2.6x    & 14.6 & 2.5x    & 68.2  & 2.5x \tabularnewline
            NTL\cite{shoup2001ntl}          & 2.12 & 3.8x    & 9.05 & 4.1x    & 42.4  & 4.1x \tabularnewline
            \intel{} AVX512-IFMA52          & 1.23 & 6.7x    & 5.72 & 6.6x    & 32.4  & 5.3x \tabularnewline
			\bottomrule
		\end{tabular}
	\end{center}
\end{table}

\subsubsection{Polynomial Kernels}
We benchmark the performance of the element-wise vector-vector and vector-scalar modular multiplication kernels.
For element-wise vector-vector modular multiplication, we compare three implementations: 1) the native C++ integer implementation; 2) the \intel{} AVX512-DQ integer implementation; 3) the \intel{} AVX512-DQ floating-point implementation.
As with the NTT, we consider three input sizes: $N=1024, 4096, 16384$.

Table~\ref{tab:results:eltwise_mod_mul} shows the runtimes for the element-wise vector-vector modular multiplication.
The \intel{} AVX512-DQ integer implementation provides a 1.5x\textendash{}1.9x speedup over the native implementation, which increases to 5.1x\textendash{}6.0x with the \intel{} AVX512-DQ floating-point implementation.

\begin{table}
	\begin{center}
        \ifiacr
        \else
		\small
		\renewcommand\theadfont{\small}
        \fi
		\caption{Single-threaded, single-core runtime in microseconds of element-wise vector-vector modular multiplication with \texttt{input\_mod\_factor} = 1.}
		\label{tab:results:eltwise_mod_mul}
		\begin{tabular}{l l l l l l l}
            \toprule
            \multirow{2}{*}{Implementation} &
			\multicolumn{6}{c}{{$N$ / Speedup}} \\
            \cmidrule(lr){2-7}
                                     & 1024   &      & 4096  &      & 16384 &       \tabularnewline
            \midrule
            Native C++ Int           & 1.51   & 1.0x & 5.71  & 1.0x & 23.6  & 1.0x  \tabularnewline
            \intel{} AVX512-DQ Int   & 0.982  & 1.5x & 3.43  & 1.6x & 12.3  & 1.9x  \tabularnewline
            \intel{} AVX512-DQ Float & 0.251  & 6.0x & 1.08  & 5.2x & 4.58  & 5.1x  \tabularnewline
			\bottomrule
		\end{tabular}
	\end{center}
\end{table}

For the element-wise vector-scalar modular multiplication kernel, we compare three implementations: 1) the native C++ implementation; 2) the \intel{} AVX512-DQ implementation; 3) the \intel{} AVX512-IFMA52 implementation.

Table~\ref{tab:results:eltwise_fma_mod} shows the runtimes for the element-wise vector-scalar modular multiplication with scalar addition.
The \intel{} AVX512-DQ implementation no significant speedup over the native implementation, as the compiler's auto-vectorizer does a sufficient job using \intel{} AVX instructions.
The \intel{} AVX512-IFMA52 implementation provides a moderate 1.7x speedup over the native implementation.

\begin{table}
	\begin{center}
        \ifiacr
        \else
		\small
		\renewcommand\theadfont{\small}
        \fi
		\caption{Single-threaded, single-core runtime in microseconds of element-wise vector-scalar modular multiplication with scalar addition and \texttt{input\_mod\_factor} = 1. }
		\label{tab:results:eltwise_fma_mod}
		\begin{tabular}{l l l l l l l}
            \toprule
            \multirow{2}{*}{Implementation} &
			\multicolumn{6}{c}{{$N$ / Speedup}} \\
            \cmidrule(lr){2-7}
                                   & 1024  &       & 4096 &       & 16384 &     \tabularnewline
            \midrule
            Native C++             & 0.53  & 1.0x  & 2.11 & 1.0x  & 9.01 & 1.0x \tabularnewline
            \intel{} AVX512-DQ     & 0.53  & 1.0x  & 2.11 & 1.0x  & 9.01 & 1.0x \tabularnewline
            \intel{} AVX512-IFMA52 & 0.302 & 1.7x  & 1.20 & 1.7x  & 5.08 & 1.7x \tabularnewline
			\bottomrule
		\end{tabular}
	\end{center}
\end{table}

\subsection{HE Kernels}
We benchmark the performance of two HE libraries that have adopted \projectname{}, Microsoft SEAL~\cite{sealcrypto} and PALISADE~\cite{palisade}.
We benchmark Microsoft SEAL~\cite{sealcrypto} version 3.6.5 and PALISADE~\cite{palisade} version 1.11.3.
For each library, we compile with and without \projectname{} support and compare the throughput.
We compile PALISADE using \emph{WITH\_OPENMP=OFF} and \emph{WITH\_NATIVEOPT=ON} for best single-threaded performance.

Table~\ref{tab:results:seal} and Table~\ref{tab:results:palisade} show the runtime speedup in Microsoft SEAL~\cite{sealcrypto} and PALISADE~\cite{palisade} HE kernels due to \projectname{}.
The amount of speedup for each kernel depends on several factors, including which \projectname{} functions the implementation calls for the underlying computation and how efficiently the HE library implements each kernel natively.
We report only the speedup rather than the execution time of each kernel to avoid any performance comparison between Microsoft SEAL~\cite{sealcrypto} and PALISADE~\cite{palisade}.
There are several implementation details that differ between the libraries that make such a comparison tricky beyond the scope of this work.
Rather, the main point here is that \projectname{} has a flexible enough API to support different HE libraries, and that the integration at the polynomial level is effective in improving overall performance of the library.

\begin{table}
	\begin{center}
        \ifiacr
        \else
		\small
		\renewcommand\theadfont{\small}
        \fi
		\caption{Speedup in microseconds of single-threaded, single-core Microsoft SEAL~\cite{sealcrypto} benchmarks with polynomial modulus degree $N=8192$ and $L=3$ 50-bit coefficient moduli (plus one auxiliary prime used for relinearization only).}
		\label{tab:results:seal}
		\begin{tabular}{l l }
            \toprule
            Benchmark & Speedup    \\
            \midrule
            Forward NTT               & 4.70x \tabularnewline
            Inverse NTT               & 5.46x \tabularnewline
            BFV Encrypt               & 1.54x \tabularnewline
            BFV Decrypt               & 2.48x  \tabularnewline
            BFV Multiply              & 1.43x \tabularnewline
            BFV Multiply Relinearize  & 1.56x \tabularnewline
            BFV Rotate 1              & 2.38x  \tabularnewline
            CKKS Encode               & 1.68x \tabularnewline
            CKKS Decode               & 1.23x \tabularnewline
            CKKS Encrypt              & 1.87x \tabularnewline
            CKKS Decrypt              & 2.80x \tabularnewline
            CKKS Multiply             & 2.66x \tabularnewline
            CKKS Multiply Relinearize & 2.22x \tabularnewline
            CKKS Rescale              & 3.26x \tabularnewline
            CKKS Rotate 1             & 2.08x \tabularnewline
			\bottomrule
		\end{tabular}
	\end{center}
\end{table}

\begin{table}
	\begin{center}
        \ifiacr
        \else
		\small
		\renewcommand\theadfont{\small}
        \fi
		\caption{Speedup in microseconds of single-threaded, single-core PALISADE~\cite{palisade} benchmarks with polynomial modulus degree $N=8192$ and $L=3$ 50-bit coefficient moduli.}
		\label{tab:results:palisade}
		\begin{tabular}{l l l l}
            \toprule
            Benchmark & Speedup    \\
            \midrule
            Forward NTT                       & 6.26x \tabularnewline
            Inverse NTT                       & 4.83x \tabularnewline
            BFV Encode                        & 2.84x \tabularnewline
            BFV Decode                        & 1.72x \tabularnewline
            BFV Encrypt                       & 1.23x \tabularnewline
            BFV Decrypt                       & 1.91x \tabularnewline
            BFV Multiply                      & 1.50x  \tabularnewline
            BFV Rotate 1                      & 2.12x \tabularnewline
            CKKS Encode                       & 1.69x \tabularnewline
            CKKS Encrypt                      & 1.19x \tabularnewline
            CKKS Multiply                     & 2.59x \tabularnewline
            CKKS Multiply Relinearize Rescale & 3.24x \tabularnewline
            CKKS Rotate 1                     & 2.68x \tabularnewline
			\bottomrule
		\end{tabular}
	\end{center}
\end{table}

\section{Conclusion}  \label{sec:conclusion}

Here, we introduced \projectname{}, a C++ library using the \intel{} AVX512 instruction set to accelerate key primitives in lattice cryptography.
\projectname{} provides optimized implementations of the number-theoretic transform (NTT) and polynomial operations, including element-wise vector-vector modular multiplication and element-wise vector-scalar modular multiplication.
The \intel{} AVX512-DQ instruction set is used to accelerate the operations for a wide range of word-sized primes, up to 62 bits.
The recent \intel{} AVX512-IFMA52 extension to the \intel{} AVX512 instruction set further improves performance for primes less than 50\textendash{}52 bits.
In particular, the \intel{} AVX512-IFMA52 instructions yield up to 7.2x and 6.7x single-threaded speedup over a native C++ implementation of the forward and inverse NTT, respectively.
The \intel{} AVX512-DQ floating-point implementation of element-wise modular multiplication yields up to 6.0x single-threaded speedup over the native C++ implementation, while the \intel{} AVX512-IFMA52 implementation of element-wise vector-scalar modular multiplication yields 1.7x single-threaded speedup over the native C++ implementation.
The \projectname{} library is available open-source at \repourl{} under the Apache 2.0 license.
\projectname{} has been adopted by the Microsoft SEAL~\cite{sealcrypto} and PALISADE~\cite{palisade} homomorphic encryption libraries.

Future work improving \projectname{} includes exploring additional NTT implementations, such as higher-radix implementations.
In particular, higher-radix NTT implementations may reduce the memory pressure, which currently bottlenecks the larger-size NTT performance, as observed in~\cite{jung2020heaan}.
We also plan to integrate \projectname{} with additional open-source homomorphic encryption libraries, as well as expand the API to encompass a larger variety of applications.
Adding a programming model which enables compiling several \projectname{} kernels in sequence may enable cross-kernel compiler optimizations such as loop fusion for additional performance improvement.

\section*{Acknowledgements}

We would like to thank Ilya Albrekht for guidance on the AVX512 implementation.
We would also like to thank Kim Laine and  Wei Dai for their support integrating \projectname{} to Microsoft SEAL~\cite{sealcrypto} as well as Kurt Rohloff and Yuriy Polyakov for their support integrating \projectname{} to PALISADE~\cite{palisade}.